\documentclass[a4paper]{article}

\usepackage{INTERSPEECH2022}

\PassOptionsToPackage{hyphens}{url}\usepackage{hyperref}

\usepackage{cite}
\usepackage{amsmath,amssymb,amsfonts}
\usepackage{algorithmic}
\usepackage{graphicx}
\usepackage{textcomp}
\usepackage{xcolor}
\usepackage[dvips]{epsfig}
\usepackage{tablefootnote}

\usepackage{multirow}
\usepackage{makecell}
\usepackage{color}

\usepackage{enumitem}

\newcommand{\feedit}[1]{\textcolor{black}{#1}}

\newcommand{\fredit}[1]{\textcolor{black}{#1}}

\newcommand{\reedit}[1]{\textcolor{black}{#1}}

\newcommand{\rredit}[1]{\textcolor{black}{#1}}

\def\BibTeX{{\rm B\kern-.05em{\sc i\kern-.025em b}\kern-.08em
    T\kern-.1667em\lower.7ex\hbox{E}\kern-.125emX}}

\title{TTS-by-TTS 2: Data-Selective Augmentation for Neural Speech Synthesis Using Ranking Support Vector Machine with Variational Autoencoder}
\name{Eunwoo Song$^1$, Ryuichi Yamamoto$^{2}$, Ohsung Kwon$^1$, Chan-Ho Song$^1$, Min-Jae Hwang$^1$,\\ Suhyeon Oh$^1$, Hyun-Wook Yoon$^1$, Jin-Seob Kim$^1$, Jae-Min Kim$^1$}
\address{
    $^{1}$NAVER Corp., Seongnam, Korea\\
    $^{2}$LINE Corp., Tokyo, Japan}

\begin{document}
\fontsize{8.7}{10.6}\selectfont

\maketitle
\begin{abstract}
    Recent advances in synthetic speech quality have enabled us to train text-to-speech (TTS) systems by using synthetic corpora. 
    However, merely increasing the amount of synthetic data is \feedit{not} always advantageous for improving training efficiency. 
    Our aim in this study is to selectively choose synthetic data that are beneficial to the training process. 
    In the proposed method, we first adopt a variational autoencoder whose posterior distribution is utilized to extract latent features representing acoustic similarity between the recorded and synthetic corpora. 
    By using those learned features, we then train a ranking support vector machine (RankSVM) that is well known for effectively ranking relative attributes among binary classes. 
    By setting the recorded and synthetic ones as two opposite classes, RankSVM is used to determine how the synthesized speech is acoustically similar to the recorded data. 
    Then, synthetic TTS data, \reedit{whose distribution is close} to the recorded data, are selected from large-scale synthetic corpora. 
    By using these data for \feedit{retraining} the TTS model, the \reedit{synthetic quality} can be significantly improved. 
    Objective and subjective evaluation results \reedit{show the superiority of the proposed method over the conventional methods}.
    
\end{abstract}

\section{Introduction}

    As the accuracy of acoustic modeling has increased following the revolution of deep neural networks, the synthetic quality of neural text-to-speech (TTS) systems has improved significantly \cite{yuxuan2017tacotron, jonathan2017natural, li2019close}.
    However, these systems still have a major shortcoming in that \feedit{a lot of} training \feedit{corpora are required to learn} the complex \feedit{nature} of speech production \cite{chung2019semi}.
    
    To overcome this limitation, various studies employing data augmentation techniques have been proposed.
    For instance, Huybrechts et al. \cite{huybrechts2021low, ribeiro2022cross} proposed using a well-trained voice conversion model to extend the speaking style of the target speaker's TTS acoustic model; 
    Wu et al. \cite{wu2021relational} proposed a speaker similarity-based data selection method from other speakers' recordings for enlarging the training corpus of the TTS vocoding model.
    While those methods require external datasets to augment target speakers' voices, Sharma et al. \cite{sharma2020strawnet} proposed to generate a large-scale synthetic corpus within the same speaker's model to distill the knowledge from the autoregressive (AR) WaveNet to the non-AR Parallel WaveNet.
    
    Similar to Sharma's work, our previous work proposed a TTS-by-TTS model in which a large-scale synthetic corpus generated by a well-designed TTS model is used to improve the quality of other TTS models \cite{hwang2021tts}. 
    One of the key ideas of this work was to collect a large number of text scripts while maintaining the recording scripts' phoneme distribution. 
    This enabled the model to simulate various phoneme combinations, resulting in significantly improving the TTS model's stability with the unseen text. 
    
    However, increasing the amount of the corresponding synthetic waveform is \feedit{not} always advantageous for improving training efficiency \cite{oh2022effective}, since this might cause \feedit{negative} effects \feedit{if} poorly generated \feedit{waveforms} are included. 
    To address this problem, we propose a data-selective augmentation method for TTS systems. 
    From a large-scale synthetic corpus, the proposed method can selectively choose the training data whose acoustic distribution is similar to the recordings. 
    \feedit{Specifically}, we adopt a \textit{variational autoencoder} (VAE) as a reference encoder for the TTS acoustic model \cite{kingma2014auto}. 
    VAEs are well known for capturing latent representations of feature distribution and have been employed as style encoders of controllable and/or expressive TTS tasks \cite{akuzawa2018expressive, zhang2019learning, sun2020fully, kim2021conditional}. 
    Similarly, our method utilizes \rredit{a VAE} to capture acoustic similarities between the acoustic features extracted from the natural recordings and synthesized by the TTS models.
    
    We introduce a \textit{ranking support vector machine} (RankSVM) \cite{parikh2011relative}, which is well known for scoring relative attributes between binary classes. 
    \feedit{Diverging from our previous work \cite{oh2022effective} that used OpenSMILE features \cite{eyben2010opensmile}, we employ} \fredit{latent} feature vectors extracted from the learned VAE's posterior \rredit{distribution} for recorded and synthetic ones.    
    By setting \feedit{them} as two opposite classes, the RankSVM learns to determine \textit{originalities} that represent how the distribution of the synthetic waveform is acoustically \feedit{close} to that of the natural recordings.
    On the basis of this originality, it is possible to selectively discard the synthetic \feedit{waveforms whose attributes are} dissimilar to the recordings; therefore, the modeling accuracy of the TTS system retrained with the remaining samples becomes significantly improved. 

    The objective and subjective evaluation results verified the \feedit{superiority of the proposed method over} the original system trained with recorded data alone and the similarly configured system retrained with all the synthetic data without any selection method.
    \reedit{In particular, our method achieved 3.89 and 3.74 mean opinion score (MOS) for Tacotron~2 and FastSpeech~2 models, respectively, with 1,000 utterance\rredit{s} of limited training data.}    
    
\section{TTS-driven data augmentation}
\label{sec:txt}

   	\begin{figure}[!t]
   	\begin{minipage}[b]{1.0\linewidth}
   		\centerline{\epsfig{figure=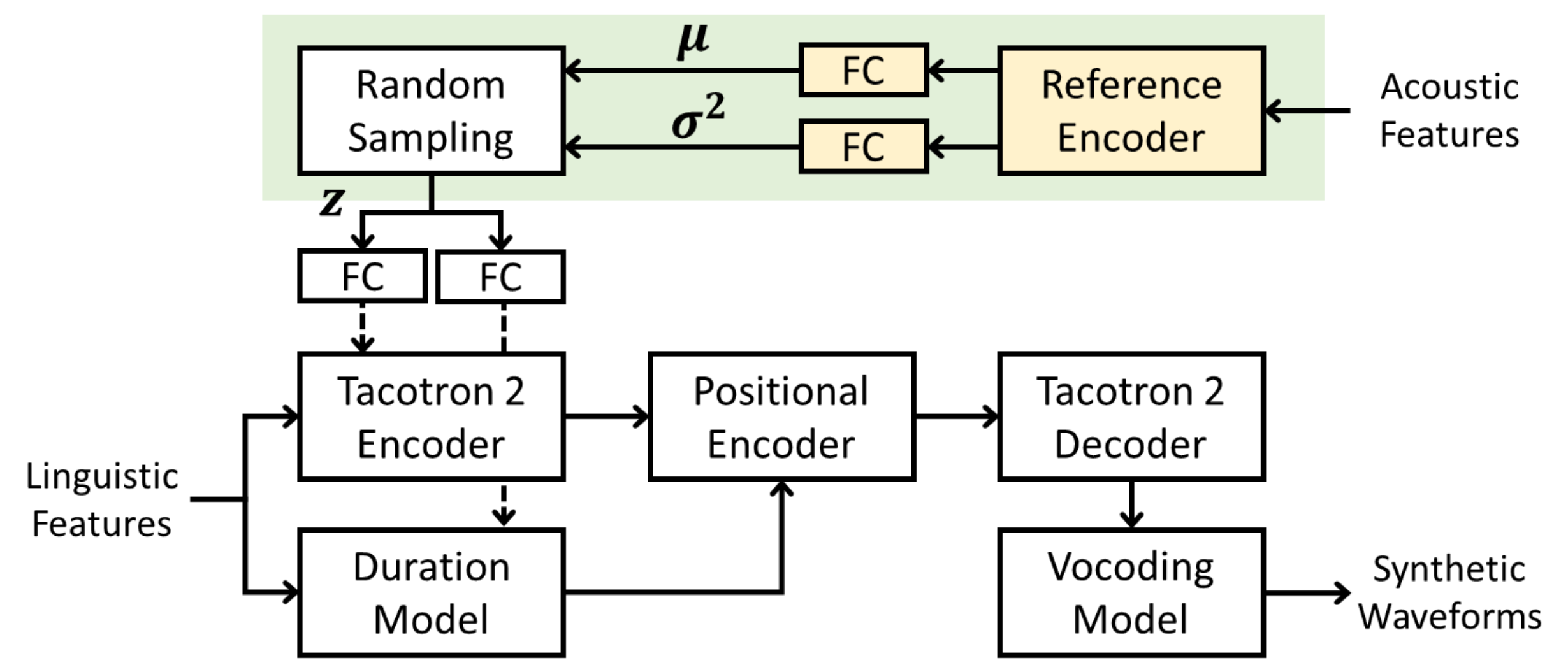,width=82mm}}
   		\centerline{(a)}  \medskip
   	\end{minipage}
   	\begin{minipage}[b]{1.0\linewidth}
   		\centerline{\epsfig{figure=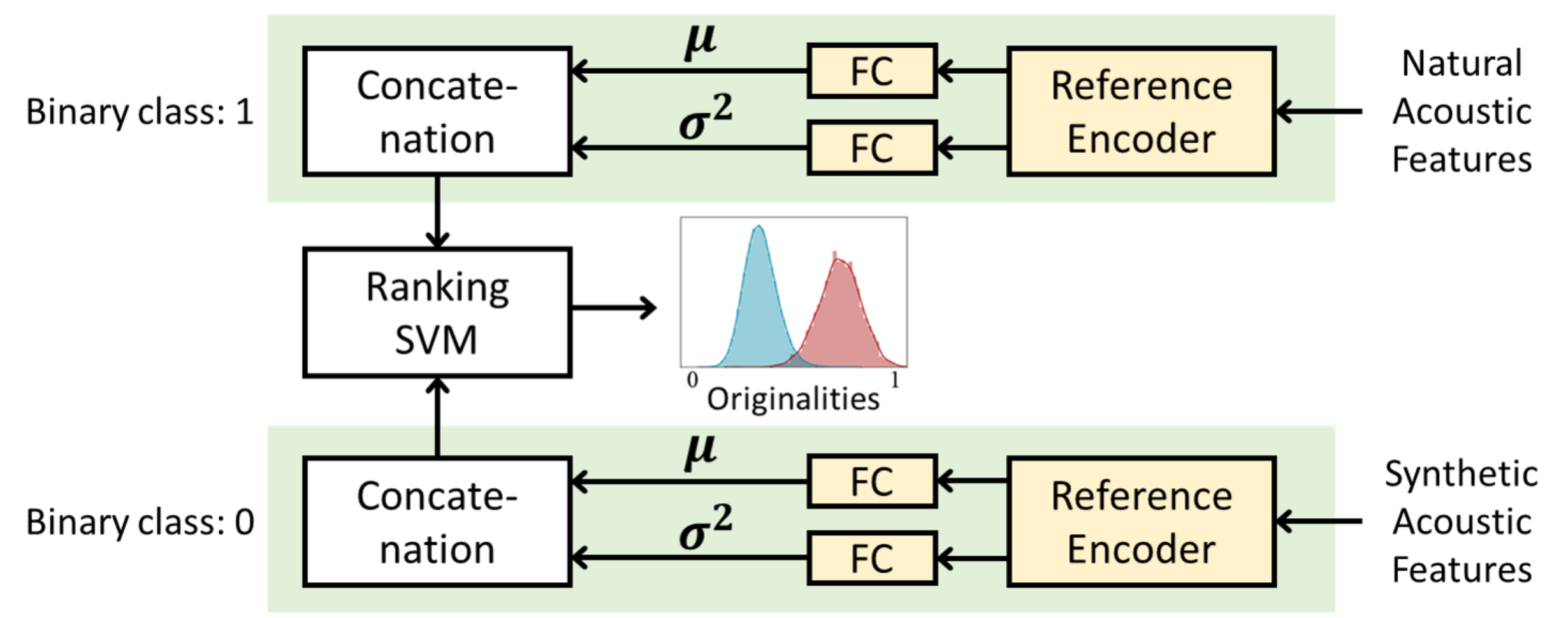,width=80mm}}
   		\centerline{(b)}  \medskip
   	\end{minipage}
    \vspace*{-15pt}  
    \caption{Block diagrams of (a) the baseline TTS model and (b) the RankSVM model. The symbols $\mu$ and $\sigma^2$ denote the VAE statistics  (i.e., the mean and variance of the posterior distribution, respectively) to sample the latent variable $z$.}    
   	\label{fig:block} 
   	\end{figure}

\subsection{Database}
\label{ssec:data}
	The experiments used a phonetically balanced Korean corpus recorded by a Korean female professional speaker. 
	The speech signals were sampled at 24 kHz with 16-bit quantization.
	In total, 1,000 (1.8 hours), 270 (0.5 hours), 130 (0.2 hours) utterances were used for training, validation, and testing, respectively. 
	
\subsection{Baseline TTS model}
\label{ssec:baseline}

    It is \feedit{crucial} to design \feedit{a} well-structured TTS \feedit{system} to synthesize high-quality speech \feedit{database}. 
    Among various state-of-the-art models, we opted to pursue a VAE-Tacotron~2 model with a phoneme alignment approach \reedit{(Figure~\ref{fig:block}a)} thanks to its stable generation and competitive synthetic quality \cite{okamoto2019tacotron, song2020neural}.
    Following the previous study \reedit{in Zhang et al.} \cite{zhang2019learning}, the prior and posterior distributions of the VAE are modeled by Gaussian distributions. 
    Note that the use of VAE is beneficial not only for giving variations to the synthetic sample itself but also for capturing the acoustic characteristics of the natural recordings and the synthetic waveforms. 
    Specifically, we leverage the statistics of the learned VAE's posterior distribution (i.e., mean and variance) to measure acoustic similarities, which will be discussed in Section~\ref{sec:proposed}.
    
    In the case of the vocoding models, we adopt a Parallel WaveGAN vocoder \cite{ yamamoto2020parallel, song2021improved, yamamoto2021parallel} \feedit{based on a} multi-band harmonic-plus-noise \feedit{model} (MB-HN-PWG \cite{hwang2021high}).
    \feedit{Specifically}, two separate WaveNet \feedit{generators} are trained to jointly \feedit{learn} the harmonic and noise \feedit{characteristics} of the speech, respectively. 
    \feedit{By employing} the target speech's periodicity in the multiple frequency bands, it is \feedit{possible to produce} speech \feedit{outputs} as naturally as recordings.
    
\subsection{\feedit{Large-scale} data augmentation}
\label{ssec:augmentation}    

    To \feedit{collect} text scripts for \feedit{synthesizing the large-scale speech waveforms}, we crawled news articles \feedit{from} the NAVER website\footnote{\url{https://news.naver.com}} \cite{hwang2021tts, oh2022effective} and \feedit{prepared} 80,000 text scripts that were 80 times larger than the training utterances.
    Using \feedit{these} text scripts, the pretrained TTS model described in the previous section generates\footnote{
        In the generation step, we condition the TTS acoustic model on VAE by using the centroid of latent vectors computed over all the training data \cite{um2020emotional}.}
    \feedit{the corresponding 80,000} speech waveforms.

\section{RankSVM-based data selection}
\label{sec:proposed}

    Increasing the number of text scripts is beneficial for the target TTS model to learn the distribution of various phoneme combinations, enabling the model to generate more stable speech from the unseen text input \cite{hwang2021tts}. 
    However, we must carefully increase the number of corresponding synthetic waveforms to exclude poorly synthesized samples before retraining the target TTS model \cite{oh2022effective}. 
    To achieve this, we propose a RankSVM-based data selection method composed of three sub-steps: VAE fine-tuning, RankSVM training, and data selection.
    
   	\begin{figure}[!t]
   	\begin{minipage}[b]{.49\linewidth}
   		\centerline{\epsfig{figure=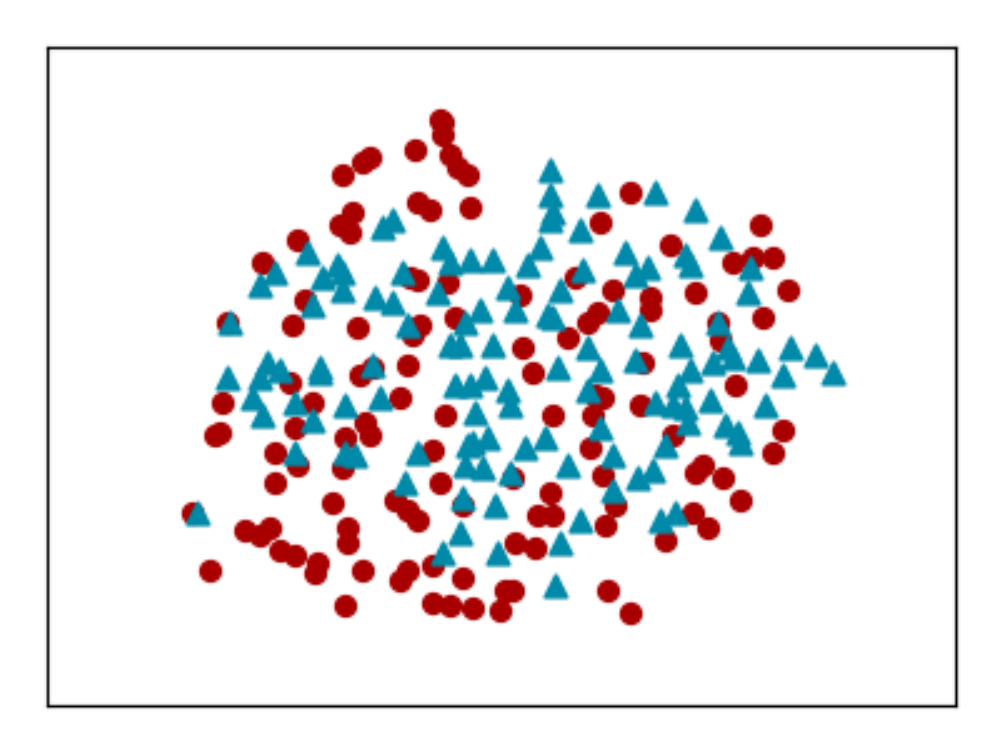,width=39mm}}
        \vspace*{-5pt}  
   		\centerline{(a)}  \medskip
   	\end{minipage}
   	\begin{minipage}[b]{.49\linewidth}
   		\centerline{\epsfig{figure=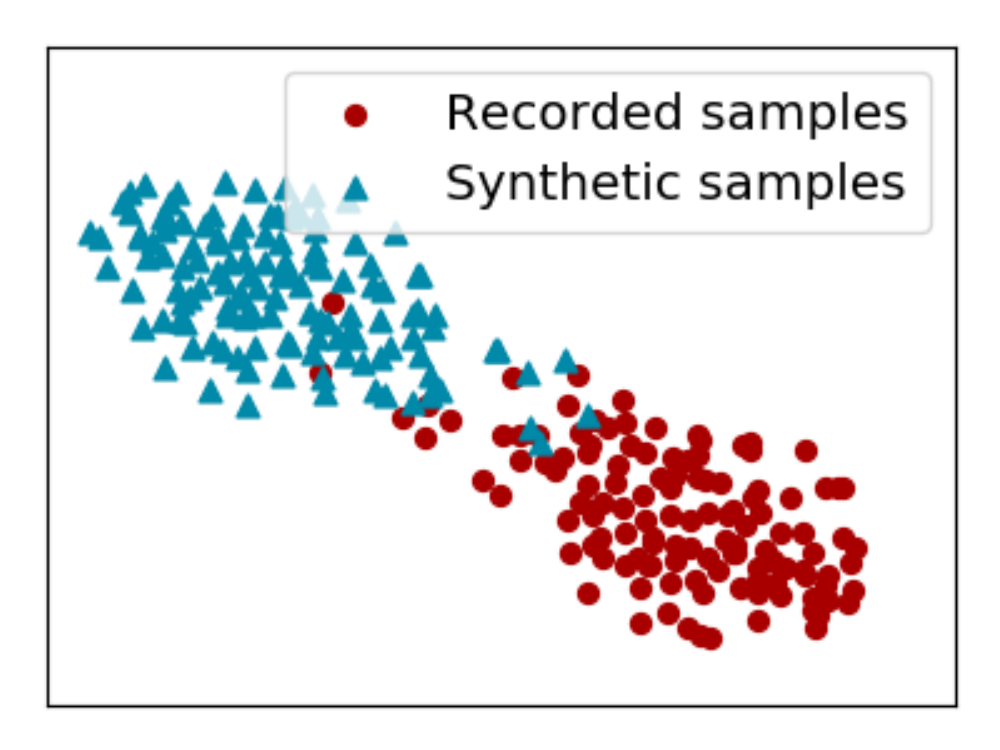,width=39mm}}
        \vspace*{-5pt}  
   		\centerline{(b)}  \medskip
   	\end{minipage}
    \vspace*{-7pt}  
    \caption{Distribution of \fredit{the} VAE's posterior mean vector extracted from recorded (red dot) and synthetic (blue triangle) samples: \rredit{(a)} before and \rredit{(b)} after applying the VAE fine-tuning process.}
    \label{fig:vae} 
    \vspace*{-6pt} 
   	\end{figure}

\subsection{VAE fine-tuning}
\label{ssec:vae}

    To effectively train the RankSVM, it is very important to extract VAE statistics that adequately represent the distributions between the recorded and synthetic waveforms. 
    However, the pretrained TTS model \rredit{is} learned through the recorded database only, and it is difficult for the VAE model to capture the synthetic database's distribution.
    
    To address this problem, we fine-tune the pretrained TTS model by using the TTS corpus combined with both recorded and synthetic corpora. 
    Because both recorded and synthetic acoustic features are now included for training, the fine-tuned VAE model becomes aware of their acoustic distribution\footnote{
        The fine-tuned VAE is only used as a feature extractor of the RankSVM model. The large-scale data augmentation described in Section~\ref{sec:txt} is performed by the pretrained model without the fine-tuning process.}. 
    Figure~\ref{fig:vae} describes the VAE's statistics approximated in the t-distributed stochastic neighbor embedding (t-SNE) space extracted from the test set \cite{van2008visualizing}. 
    It shows that applying the fine-tuning process (Figure~\ref{fig:vae}b) represents the feature distributions better, which is expected to be beneficial for training the RankSVM model.

    \begin{figure}[!t]
        \centering
		\includegraphics[width=0.7\linewidth]{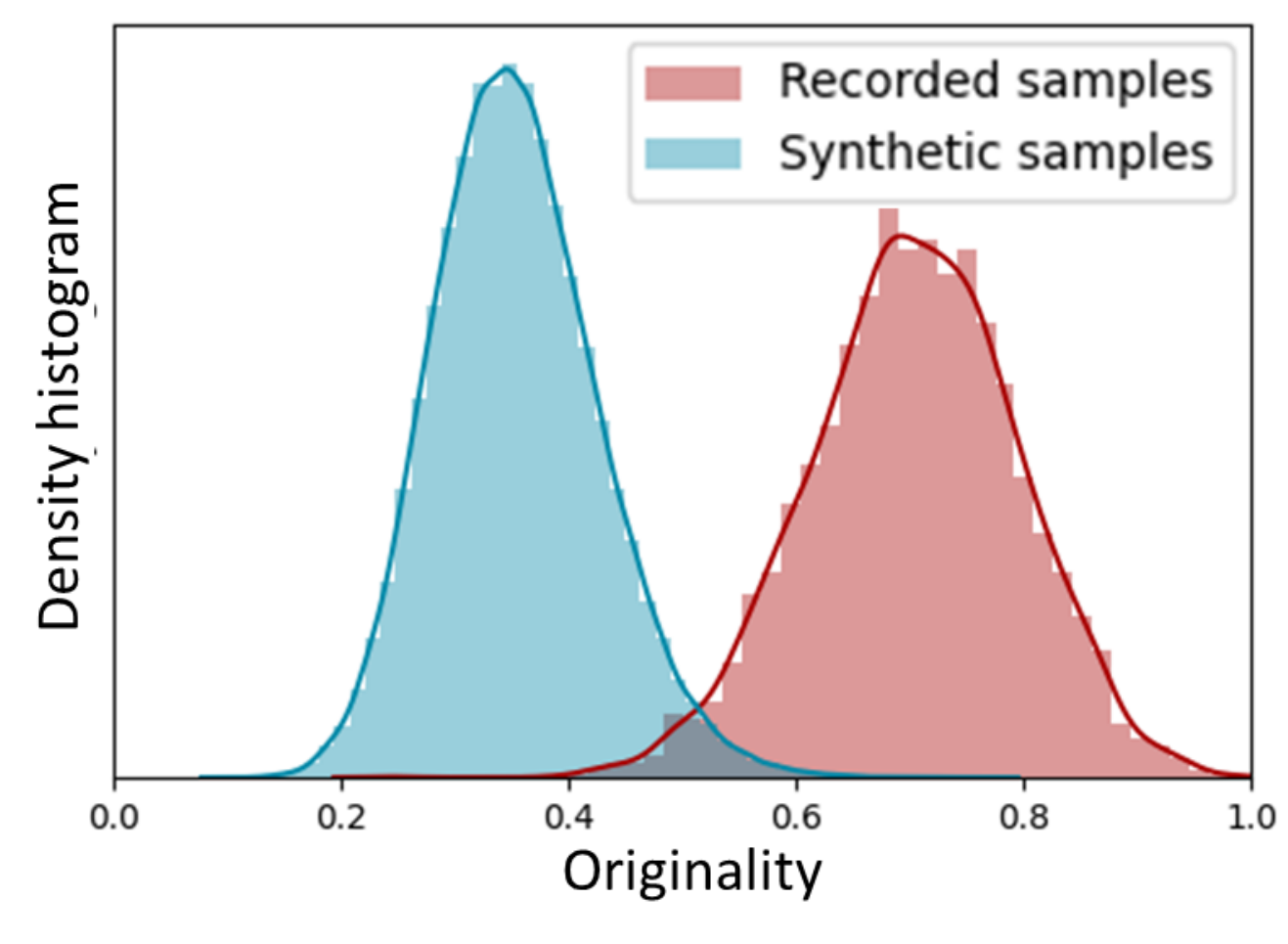}
    	\vspace*{-7pt}  
        \caption{\small
            Density histograms of originality determined using RankSVM from the recorded (red boxes) and synthetic (blue boxes) waveforms, respectively.
        }
        \label{fig:histogram}
    \vspace*{-6pt}  
    \end{figure}

\subsection{RankSVM}
\label{ssec:rankingsvm}
    RankSVM is well known for learning relative attributes from two different classes \cite{parikh2011relative}.
    As shown in Figure~\ref{fig:block}b, the proposed method sets the VAE feature vectors obtained from recorded and synthetic ones as two opposite classes, respectively. 
    Then, \feedit{a} ranking function \feedit{is learned} to \feedit{determine} the \textit{originality}, \feedit{a} real-valued score defined as how the distribution of the synthetic \fredit{waveforms} is acoustically similar to that of the natural recordings.

\subsubsection{Ranking function}
    Assuming we have a training set $T = \{t\}$ represented in $\mathbb{R}^n$ by VAE features $\{\bm{x}_t\}$, and $T = N \cup M$, where $N$ and $M$ denote the recorded and synthetic set, respectively.
    The goal of RankSVM is to learn a ranking function defined as follows \cite{deng2014linear}: 
    \begin{align}
    r(\bm{x}_t) = \bm{w}^{T}\bm{\bm{x}_t}, \label{eq:rankingfunction}
    \end{align}
    with the following constraints:
    \begin{align}
    \forall (i,j) \in O &:  \bm{w}^{T}\bm{\bm{x}_i} > \bm{w}^{T}\bm{\bm{x}_j}, \\
    \forall (i,j) \in S &:  \bm{w}^{T}\bm{\bm{x}_i} = \bm{w}^{T}\bm{\bm{x}_j},
    \end{align}
    where $\bm{w}$, $O$, and $S$ denote the weight vector of the ranking function, ordered set, and similar paired set, respectively. Specifically, the ordered set represents pairs of recorded and synthetic samples, and the similar set represents either pairs of synthetic samples or pairs of recorded samples.

    Parikh et al. \cite{parikh2011relative} proposed to solve the problem using Newton's method \cite{chapelle2007training}. 
    However, this is not practical for large-scale data augmentation because of its quadratic computational complexity with regard to the number of training samples. Therefore, we adapted a stochastic gradient decent to accelerate the training \cite{shalev2011pegasos}. 
    Once we obtain an optimal $\bm{w}$, the originality of \rredit{recorded and synthetic samples} is calculated by Equation (\ref{eq:rankingfunction}). 
    Note that the originality scores are normalized to $[0, 1]$ for convenience.
    
	\begin{table}[!t]
	\begin{center}
	\caption{
	    Distortions with a 95\% confidence interval measured from two different groups: synthetic samples that have high 10\% and low 10\% originalities, respectively.
    }
	\label{table:analysis}
	{\small        
	\begin{tabular}{ccc}
	\Xhline{2\arrayrulewidth}
	Range of originality & F0 RMSE (Hz) & LSD (dB)
    \\ \hline
	High 10\%   & 26.66$\pm$5.18 & 3.87$\pm$0.18   \\	
	Low 10\%    & 32.28$\pm$5.96 & 4.01$\pm$0.16   \\
	\Xhline{2\arrayrulewidth}
	\end{tabular}}	
	\end{center}   
    \vspace*{-12pt}
	\end{table}
    
\subsubsection{Relationship between originality and acoustic similarity}
    
    To further verify the relationship between the synthetic sample's originality and its acoustic similarity with the natural recordings, we analyzed distortions in the generated acoustic features. 
    In detail, we collected two different synthetic groups from the test set. 
    One group consisted of synthetic samples that had high 10\% originalities, whereas the other group consisted of synthetic samples that had low 10\% originalities. Then, objective metrics, such as F0 root mean square error (RMSE) and log spectral distance (LSD), were measured from each group.
    
    Table~\ref{table:analysis} shows the test results. 
    It can be seen that the first group has much smaller generation distortions, which confirms that the synthetic waveforms with high originality have acoustically close characteristics to the natural recordings.
    
\subsection{Data selection}
\label{ssec:selection}

    On the basis of the synthetic waveform's originality, we selectively discard the samples (e.g., those with low originality) that have dissimilar characteristics with the recordings. 
    The entire acoustic model described in Section~\ref{ssec:baseline} is then retrained by using the remaining samples (e.g., those with high originality) together with the original recordings. 
    Note that the proposed task is not applied to the vocoding model since it has been reported that using large sizes of training data is not \feedit{important} for \feedit{training the vocoding model} \cite{okamoto2020realtime}. 
    \feedit{Consistently}, our previous work \feedit{also} confirmed \feedit{that including the} augmented database \feedit{to the training process was not that effective to improve the vocoding} quality \cite{hwang2021tts}. 
    
\section{Experiments}
\label{sec:experiment}

\subsection{TTS model details}
\subsubsection{Feature extraction}
    The linguistic features \feedit{consisted of} 354-dimensional feature vectors \feedit{containing} 330 binary features for categorical contexts and 24 \feedit{additional} features for numerical contexts. 
    The corresponding acoustic features were extracted using the improved time-frequency trajectory excitation vocoder at \feedit{every} 5 ms frame intervals \cite{song2017effective}.
    \feedit{Note that the feature dimension was 79 containing} 40-dimensional line spectral frequencies, F0, gain, binary voicing flag (v/uv), a 32-dimensional slowly evolving waveform, and a 4-dimensional rapidly evolving waveform.
    
\subsubsection{Acoustic model}

    We adopted a VAE-Tacotron~2 model with a phoneme alignment approach \cite{zhang2019learning, okamoto2019tacotron}. 
    As shown in Figure~\ref{fig:block}a, the reference acoustic features were first fed into the VAE reference encoder composed of six convolutional layers followed by a \rredit{gated recurrent unit} layer \cite{wang2018style}.    
    Its output vector was then passed through two separate fully connected (FC) layers to generate the VAE \feedit{mean and variance, respectively}. 
    Finally, the latent variables\footnote{
        They were simply added to the final layers of the duration model and the Tacotron~2 encoder, respectively. Note that two separated FC layers were used for projecting latent variables to each model.}
    were sampled via the reparameterization method \rredit{\cite{kingma2014auto}}. 

    Having input linguistic features, the duration model predicted phoneme-level duration through three FC layers followed by a unidirectional \rredit{long short-term memory (LSTM)} network.    
    The encoder transformed the same linguistic features into high-level context vectors and these were upsampled to the frame \feedit{resolution}. 
    \feedit{Note that} the encoder \feedit{consisted of} three convolution layers, a \feedit{single} bi-directional LSTM network, and a single FC layer. 
    
    The decoder was used to generate the output acoustic features. 
    First, the previously generated acoustic features were fed into the PreNet \cite{jonathan2017natural}, and those features and the context vectors from the encoder were then passed through two unidirectional LSTM layers, followed by two projection layers. 
    Finally, residual elements of the generated acoustic features \feedit{were passed through the} PostNet \cite{jonathan2017natural} to improve generation accuracy.
    
    \feedit{Both} the input \feedit{linguistic} and output \feedit{acoustic} features were normalized to have zero mean and unit variance.
    \feedit{Xavier initializer \cite{xavier2010init} and RAdam optimizer \cite{liu2020radam} were used for initializing and updating the weights, respectively.}
    More setup details are given in our previous work \cite{hwang2021tts}.
    
\subsubsection{Vocoding Model}
    
    The setups for training the MB-HN-PWG model followed our previous work in \cite{hwang2021high}. 
    The harmonic WaveNet \feedit{consisted of} 20 dilated residual blocks with two exponentially increasing dilation cycles.
    \feedit{On the other hand}, the noise WaveNet \feedit{consisted of} 10 residual blocks with one exponentially increasing dilation cycle. 
    In each WaveNet, \feedit{16 band-pass filters parameterized by windowed sinc functions with 255 filter taps were applied to divide the frequency bands.}
    The model was trained for 400K steps with RAdam optimizer \cite{liu2020radam}. 
    \fredit{The discriminator weights were fixed} for the first 100K steps, and both the generator and the discriminator were jointly trained \fredit{for the rest 300K steps} \cite{hwang2021high}.

    \begin{figure}[!t]
   	\begin{minipage}[b]{.49\linewidth}
   		\centerline{\epsfig{figure=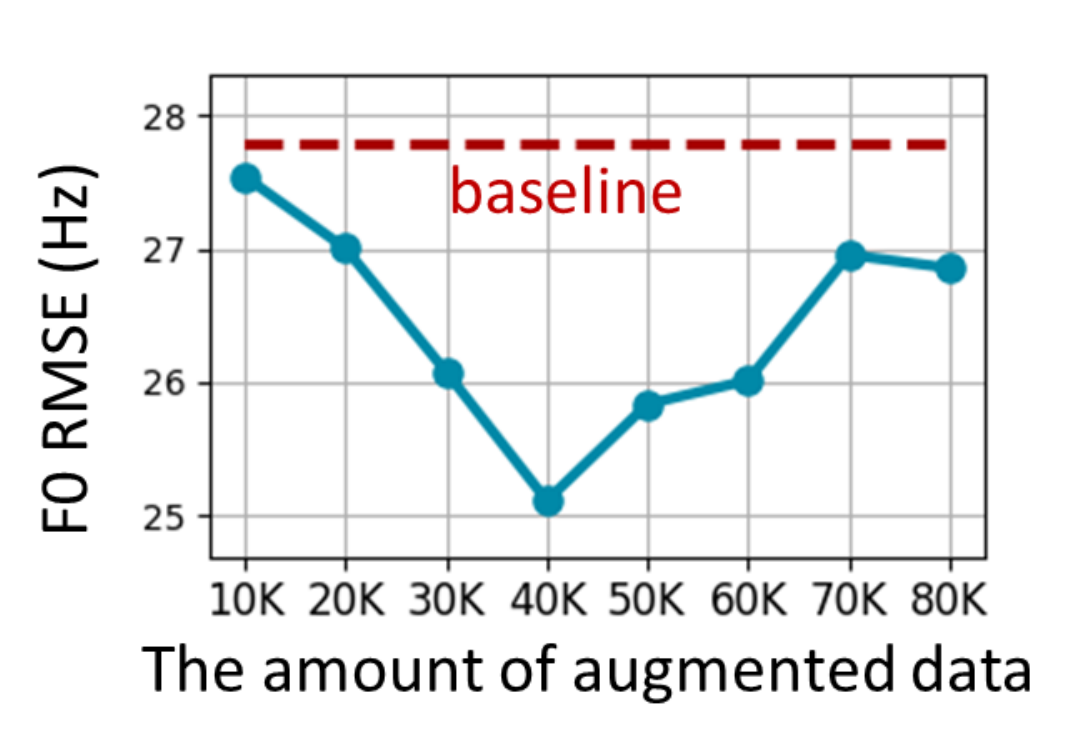,width=42mm}}
        \vspace*{-5pt}  
   		\centerline{(a)}  \medskip
        \vspace*{-5pt}  
   	\end{minipage}
   	\begin{minipage}[b]{.49\linewidth}
   		\centerline{\epsfig{figure=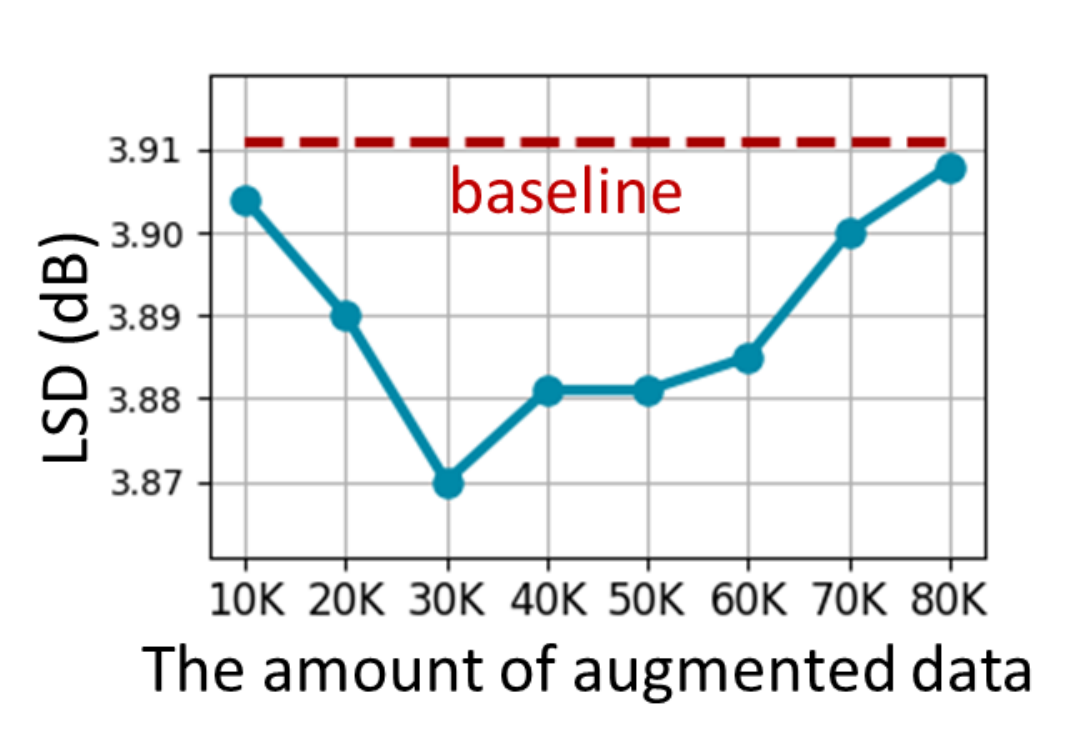,width=42mm}}
        \vspace*{-5pt}  
   		\centerline{(b)}  \medskip
        \vspace*{-5pt}  
   	\end{minipage}
  	\begin{minipage}[b]{.49\linewidth}
   		\centerline{\epsfig{figure=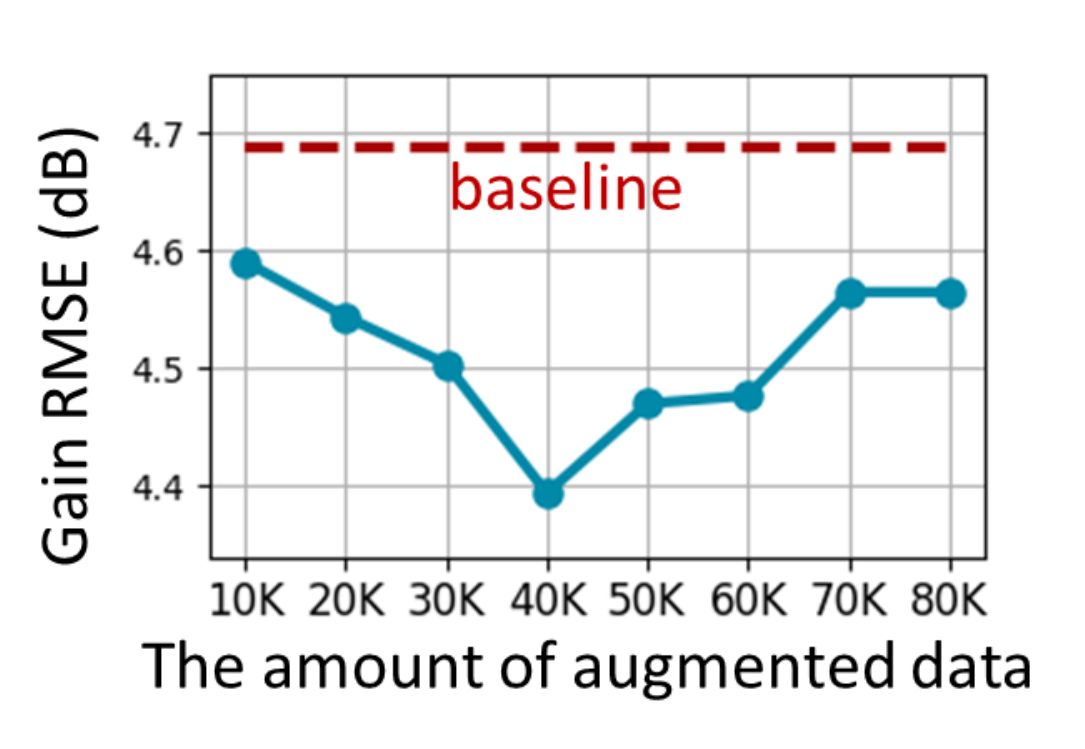,width=42mm}}
        \vspace*{-5pt}  
   		\centerline{(c)}  \medskip
   	\end{minipage}
   	\begin{minipage}[b]{.49\linewidth}
   		\centerline{\epsfig{figure=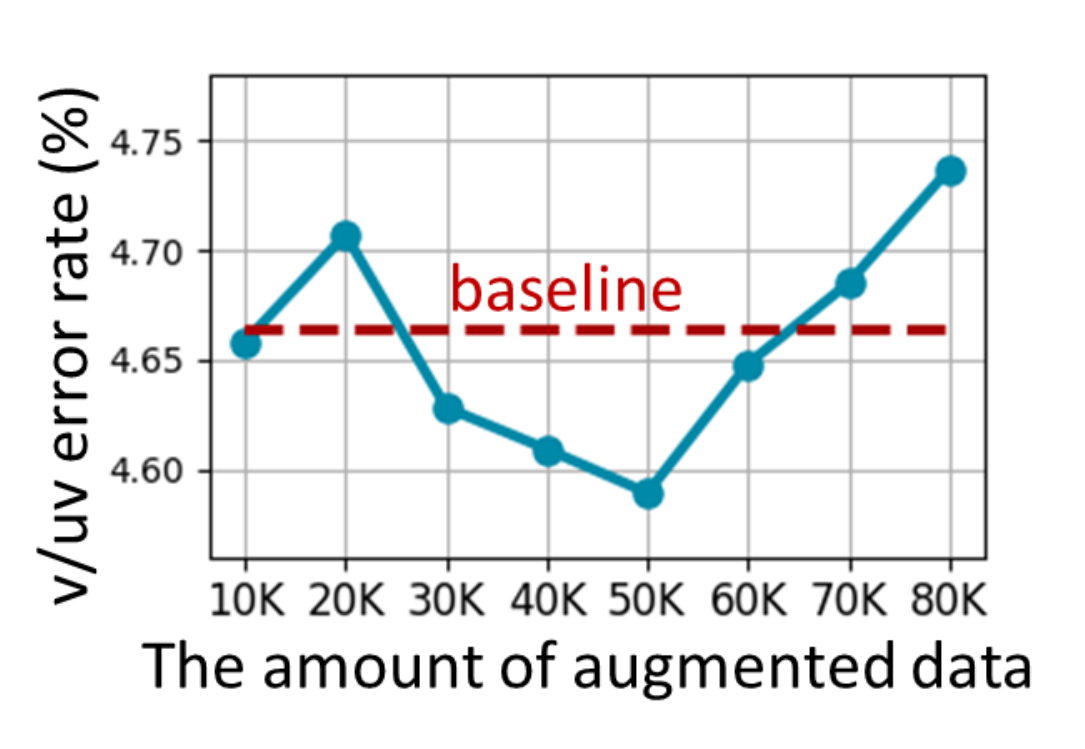,width=42mm}}
        \vspace*{-5pt}  
   		\centerline{(d)}  \medskip
   	\end{minipage}
    \vspace*{-10pt}  
    \caption{Objective evaluation results with respect to various amounts of augmented data used in the retraining process: the dashed red line represents the results of the baseline model learned with recorded data alone.}
   	\label{fig:obj} 
    \end{figure}
    \begin{figure}[!t]
        \centering
		\includegraphics[width=0.72\linewidth]{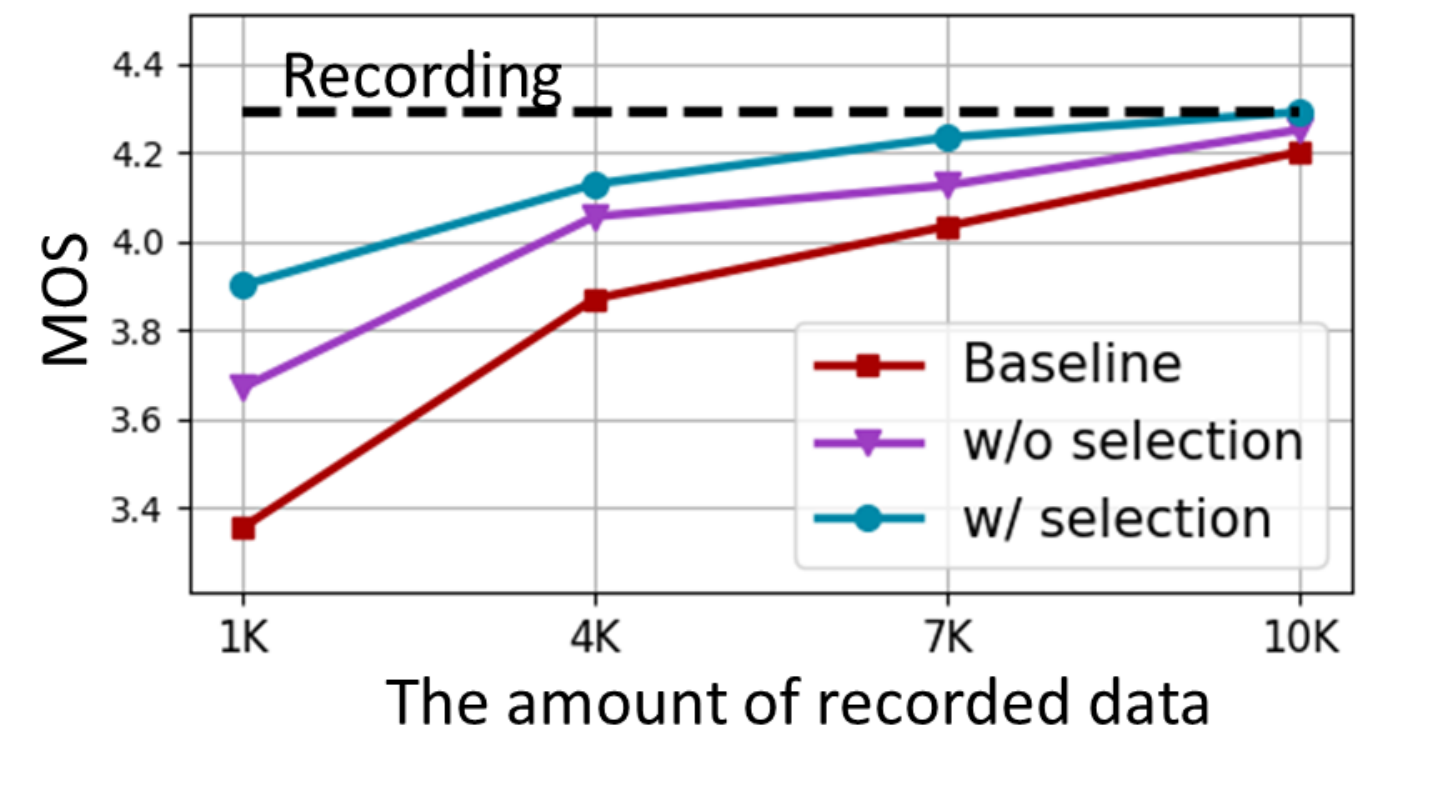}
    	\vspace{-3mm}
    	\caption{\small
            The MOS test results with respect to various amounts of recorded data: baseline model trained with recorded data alone (red square) and augmented models trained without (purple triangle) and with (blue dot) the proposed data selection method.
        }
        \label{fig:mos}
    \vspace*{-12pt}
    \end{figure}
    
\subsection{Objective \feedit{and subjective} evaluations}
\label{ssec:eval}

    To evaluate the performance of the proposed system, \feedit{we measured} the distortions \feedit{between} the acoustic \feedit{features} obtained from the original \feedit{recording} and \feedit{generated} by the TTS models.
    The metrics for measuring distortion were F0 RMSE, LSD, gain RMSE, and v/uv error rate.
    
    Figure~\ref{fig:obj} shows the objective evaluation results with respect to various amounts of augmented data sets used in the retraining process (10K, 20K, ..., 80K utterances). 
    The findings can be analyzed as follows: 
    (1) The distortions decreased when the number of augmentation databases increased to some amount (about 40K utterances) because the modeling accuracy was also improved, depending on the amount of training databases. 
    (2) However, distortions increased when the model was trained with more than 50K augmented utterances since the amount of poorly generated data increased as well. This implies the importance of data-selective augmentation in the TTS task, which will be further analyzed in the following subjective evaluations.

    To evaluate the perceptual quality of the proposed system, \feedit{we performed} five-scale naturalness MOS listening tests\footnote{
        Generated audio samples are available at the following URL:\\ \url{https://sewplay.github.io/demos/txt2/}}. 
    \feedit{In the test}, twenty native Korean listeners were asked to make quality judgments about 20 randomly selected samples from the test set.
    
    For comparison, we included the FastSpeech (FS)~2\fredit{-based TTS system} \cite{ren2020fastspeech2}. 
    The structure was similar to the Tacotron~2-based baseline TTS system, but only the acoustic model was replaced with \fredit{the} non-AR \rredit{Transformer} networks \cite{ren2019fastspeech}.    
    The detailed setup for conducting the FS~2 model followed those in our previous work \cite{hwang2021tts}.

    Table~\ref{table:subj} shows the MOS evaluation results for the TTS systems with respect to the different training conditions, and the analysis can be summarized as follows: 
    (1) In both the AR and non-AR models, the systems retrained with the synthetic corpus performed better than those trained with recorded data alone. 
    \feedit{It means} that \feedit{increasing the training data with the augmented wavefroms} was \feedit{effective} for \feedit{enhancing} the \feedit{synthetic} quality. 
    (2) Among the systems with augmentations, the proposed systems with the data-selective augmentation method (Tests 3 and 6) performed better than the systems without any selection method (Tests 2 and 5, respectively). 
    As poorly generated synthetic samples were discarded based on originality, the proposed TTS systems synthesized more natural speech. 
    (3) Consequently, our system achieved 3.89 and 3.74 MOS for the Tacotron~2 and FS~2 models, respectively, when the recorded data for training the model were limited to 1,000 utterances.

\subsection{\feedit{Additional experiments with enough recordings}}
\label{ssec:add_eval}

    To \feedit{further confirm} the effectiveness of the proposed method \feedit{under the} condition of \feedit{the enough recordings}, \feedit{we} conducted additional MOS tests by changing \textit{the number of recorded datasets} from 1K to 10K utterances. 
    In each case, we trained the baseline VAE-Tacotron~2 model, augmented 80K utterances, selected 40K utterances by using \fredit{\feedit{the} fine-turned VAE with \feedit{the} RankSVM} model, and retrained the target TTS model.     
    As shown Figure~\ref{fig:mos}, the perceptual quality was improved as the amount of recorded data increased. 
    Although the gap between the proposed and conventional methods was reduced as the synthetic quality was also improved, the data-selective augmentation method was still effective, even when the size of the source database increased enough.
    
	\begin{table}[!t]
	\begin{center}
	\caption{The MOS test results with a 95\% confidence interval: the system trained with the proposed method is shown in boldface. Note that the number of recorded utterances was 1K, and that of augmented utterances was represented as $M$.}
	\label{table:subj}
	{\small        
	\begin{tabular}{ccccc}
	\Xhline{2\arrayrulewidth}
    \multicolumn{1}{c}{\multirow{2}{*}{System}} & Model & \multicolumn{1}{c}{\multirow{2}{*}{$M$}} & Data   & \multicolumn{1}{c}{\multirow{2}{*}{MOS}}\\ 
    \multicolumn{1}{c}{}                        & type  & \multicolumn{1}{c}{}              & selection     & \multicolumn{1}{c}{} \\ \hline
	Test 1    & Tacotron 2 & -    & -   & 3.35$\pm$0.11 \\	
	Test 2    & Tacotron 2 & 80K & No  & 3.66$\pm$0.11 \\	
	\textbf{Test 3}    & \textbf{Tacotron 2} & \textbf{40K} & \textbf{Yes} &\textbf{3.89$\pm$0.09} \\
	Test 4    & FastSpeech 2 & -    & - &3.30$\pm$0.12 \\	
	Test 5    & FastSpeech 2 & 80K & No &3.61$\pm$0.09 \\	
	\textbf{Test 6}    & \textbf{FastSpeech 2} & \textbf{40K} & \textbf{Yes} &\textbf{3.74$\pm$0.08} \\	\hline 
	Recording & - & - & - & 4.29$\pm$0.08 \\	
			\Xhline{2\arrayrulewidth}
	\end{tabular}}	
	\end{center}   
    \vspace*{-18pt}
	\end{table}

\section{Conclusion}
    We proposed a TTS-driven data-selective augmentation \feedit{technique}.
    From the \feedit{large-scale synthetic} corpora, \rredit{a} RankSVM with VAE's posterior distribution determined the originality \feedit{that represents how the acoustic characteristics} of the generated speech \feedit{was similar to those of the natural recordings}.
    By selectively including the \feedit{synthetic} data \rredit{with} the \feedit{recorded one}, the performance \feedit{of the} retrained TTS system \feedit{has been improved significantly}.
    \feedit{Future works include to extend the proposed method to augment other speaking styles for the emotional and/or expressive TTS models.}
    

\section{Acknowledgment}
This work was supported by Clova Voice, NAVER Corp., Seongnam, Korea.

\bibliographystyle{IEEEtran}
\bibliography{mybib}

\end{document}